\input lanlmac.tex
\overfullrule=0pt
\input epsf.tex
\figno=0
\def\fig#1#2#3{
\par\begingroup\parindent=0pt\leftskip=1cm\rightskip=1cm\parindent=0pt
\baselineskip=11pt
\global\advance\figno by 1
\midinsert
\epsfxsize=#3
\centerline{\epsfbox{#2}}
\vskip 12pt
{\bf Fig. \the\figno:} #1\par
\endinsert\endgroup\par
}
\def\figlabel#1{\xdef#1{\the\figno}}
\def\encadremath#1{\vbox{\hrule\hbox{\vrule\kern8pt\vbox{\kern8pt
\hbox{$\displaystyle #1$}\kern8pt}
\kern8pt\vrule}\hrule}}
 
%

\def\np#1#2#3{{\it Nucl. Phys.} {\bf B#1} (#2) #3}
\def\pl#1#2#3{{\it Phys. Lett. }{\bf B#1} (#2) #3}

\def\physrev#1#2#3{{\it Phys. Rev.} {\bf D#1} (#2) #3}

\def\cmp#1#2#3{{\it Comm. Math. Phys.} {\bf #1} (#2) #3}
\def\mpl#1#2#3{{\it Mod. Phys. Lett. }{\bf #1} (#2) #3}
\def\ijmp#1#2#3{{\it Int. J. Mod. Phys.} {\bf #1} (#2) #3}
\def\lmp#1#2#3{{\it Lett. Math. Phys.} {\bf #1} (#2) #3}
\def\hepth#1{{\tt hep-th}#1}
\font\zfont = cmss10 
\font\litfont = cmr6

\def\bigone{\hbox{1\kern -.23em {\rm l}}}     
\def\ZZ{\hbox{\zfont Z\kern-.4emZ}}
\def\hf{{\litfont {1 \over 2}}}

\def\Im{{\rm Im ~}}

\def\p{\partial}

\def\a{\alpha}

\def\g{\gamma}

\def\l{\lambda}
\def\m{\mu}

\def\s{\sigma}

\def\z{\zeta }

\def\oo{\hat \omega   }

\def\o{\omega }


  \def\z{\zeta }
  
  \def\hb{\hbar\ }
  \def\bh{{1\over \hbar}}
   \def\Im{{\rm Im}}
    
  \def\CA {{\cal A}}

  \def\CD {{\cal D}}
  
  \def\CF {{\cal F}}

  \def\CL {{\cal L}}

  \def\CO {{\cal O}}

  \def\CT {{\cal T}}
  
  \def\CV {{\cal V}}
  \def\CW {{\cal W}}

  \def\CZ {{\cal Z}}

\def\hb{\hbar}
\def\bh{ {1\over \hbar}}
  \def\R{\relax{\rm I\kern-.18em R}}
  \font\cmss=cmss10 \font\cmsss=cmss10 at 7pt 
  \def\Z{\relax\ifmmode\mathchoice
  {\hbox{\cmss Z\kern-.4em Z}}{\hbox{\cmss Z\kern-.4em Z}} 
  {\lower.9pt\hbox{\cmsss Z\kern-.4em Z}}
  {\lower1.2pt\hbox{\cmsss Z\kern-.4em Z}}\else{\cmss Z\kern-.4em 
  Z}\fi} 
  \def\p{\partial}
  
  \def\11{1\!\! 1}
 
 \def\llangl{\left\langle\!\left\langle }
 \def\rrangl{\right\rangle \!\right\rangle }

   \def\hepth{ {\tt hep-th/}}
   

\lref\WITTENBH{E. Witten, ``String theory and black holes'', 
\physrev{44}{1991}{314}.}
\lref\HMP{A. Hanani, Y. Oz, R. Plesser, ``Topological 
Landau-Ginsburg formulation and Integrable structure of the 2d string 
theory'',
\cmp{173}{1995}{559}.}
\lref\MPR{G. Moore, M. Plesser, and S. Ramgoolam,
``Exact S-matrix for 2D string theory'', \hepth{9111035}.}  
\lref\MOORE{G. Moore, ``Gravitational phase transitions 
 and the sine-Gordon model", \hepth{9203061}.}  
\lref\HKK{ J. Hoppe, V. Kazakov and I. Kostov,   
``Dimensionally reduced SYM$_4$ as solvable matrix quantum 
mechanics'',
  \hepth{9907058} \np{571}{2000}{479}.}
  \lref\NKK{V. Kazakov, I. Kostov and  N.  Nekrasov, 
 ``D-particles, matrix Integrals and KP hierachy'',
       \hepth{9810035}, \np{557}{1999}{413}.}
\lref\DVV{R. Dijkgraaf, E. Verlinde and H. Verlinde, 
``String propagation in a black hole geometry'', 
\np{371}{1992}{269}.} 
\lref\JEV{A. Jevicki and D. Yoneya, \np{411}{1994}{64}. }
\lref\GRKL{D.Gross and I. Klebanov, \np{344}{1990}{475};
 \np{354}{1990}{459}.}
\lref\BULKA{D. Boulatov and V.Kazakov,  \ijmp{8}{1993}{809}. } 
\lref\FZZ{ V. Fateev, A. Zamolodchikov and Al. Zamolodchikov,
{\it unpublished}.}
\lref\FZ{Al. B. Zamolodchikov, {\it unpublished}; V.A. Fateev, \pl B 
357 (1995) 397.}
\lref\FZZ{V. Fateev, A. Zamolodchikov and Al. Zamolodchikov, {\it 
unpublished}.}
\lref\JM{M. Jimbo and T. Miwa, ``Solitons and Infinite Dimensional 
Lie 
Algebras'', {\it Publ. RIMS, Kyoto Univ.} {\bf 19}, No. 3 
(1983) 943.}
\lref\Hir{R. Hirota, Direct Method in Soliton Theory{\it Solitons}, 
Ed. by R.K. Bullogh and R.J. Caudrey, Springer, 1980.}
\lref\HK{E. 
Hsu and D. Kutasov, hep-th/9212023, Nucl. Phys. {\bf B396} (1993) 
693.}
\lref\UT{K. Ueno and K. Takasaki, ``Toda Lattice Hierarchy": 
in   `Group representations and systems of differential equations', 
{\it Adv. Stud. Pure Math.} {\bf 4} (1984) 1.}
\lref\Takasak{K. Takasaki:
{\it Adv. Stud. Pure Math.} {\bf 4} (1984) 139.}
 \lref\SAKA{S. Kakei, 
``Toda lattice hierarchy and Zamolodchikov's conjecture",
solv-int/9510006 }
\lref\PBM{A. Prudnikov, Yu. Brichkov, O. Marychev, Integrals and 
Series, Nauka , Moscow, 1981.}
\lref\POLCHINSKI{J. Polchinski, ``What is string theory'', 
{\it Lectures presented at the 1994 Les Houches Summer School 
``Fluctuating
Geometries in Statistical Mechanics and Field Theory''}, 
\hepth{9411028}.}
\lref\KLEBANOV{I. Klebanov, {\it Lectures delivered at the ICTP 
Spring School on String Theory and Quantum Gravity},
 Trieste, April 1991, \hepth{9108019}.}
 \lref\DASJP{S. Das and A. Jevicki, \mpl{5}{1990}{1639}.}
 \lref\DOUGLAS{M.  Douglas,  ``Conformal theory techniques 
in Large $N$ Yang-Mills Theory'', talk at the 1993 Carg\`ese meeting, 
\hepth{9311130}.}
\lref\KLEBwha{S. Gubster and I. Klebanov, \pl{340}{1994}{35}.}
\lref\KLEBwhb{I. Klebanov, \physrev{51}{1995}{1836}.}
\lref\MukhiImbimbo{ C. Imbimbo, S. Mukhi, \np{449}{1995}{553}.}
\lref\NTT{ T.Nakatsu, K.Takasaki, S.Tsujimaru, \np{443}{1995}{155}.}
\lref\ALZ{Al. Zamolodchikov, \np{432}{1994}{427}, \hepth{9409108}.}
\lref\BRH{H. Braden and  A. Hone, \pl{380}{1996}{296}.}
\lref\WI{H. Widom, \cmp{184}{1997}{653}.}
\lref\Takebe{T. Takebe, ``Toda lattice hierarchy and conservation 
laws'', 
\cmp{129}{1990}{129}.}
\lref\MuchiImbimbo{C. Imbimbo
and S. Mukhi, ``The topological matrix model of c=1 String",
\hepth{9505127}.}
\lref\KKK{V. Kazakov, I. Kostov and D. Kutasov, 
``A Matrix Model for the Two Dimensional Black Hole"
\hepth{0101011}}                                                           

\lref\MikeKDV{M.  Douglas,  ``Strings in less than one dimension and 
the generalized KdV hierarchies'',  \pl{238}{1990}{279}.}
\lref\zamolod{A. B. Zamolodchikov, {\it Sov. J. Nucl. Phys.} {\bf 44}
(1986) 529.}

\lref\BF{Byrd and Friedman,  {\it Handbook of Ellyptic Integrals for 
Engineers and Physicists},  Springer-Verlag, 1954.}
 \lref\AK{S. Alexandrov and V. Kazakov, 
 ``Correlators in 2D string theory with vortex condensation'',
 \hepth{0104094}.}
\lref\GGG{M. Gaudin, 
``Une famille \`a un param\`etre d'ensembles unitaires'',
{\it Nuclear Physics} {\bf 85} (1966) 545;
Travaux de Michel Gaudin, ``Mod\`eles exactement r\'esolus",
{\it Les Editions de Physique} 1995, p. 25.}
\lref\HK{E. Hsu and D. Kutasov, ``The gravitational sine-Gordon 
model",
 \hepth{9212023},  \np{396}{1993}{693}.}
\lref\bulka{D. Boulatov and V.Kazakov,  
 One-Dimensional String Theory with Vortices as Upside-Down 
 Matrix Oscillator,
\hepth{0012228}, \ijmp{8}{1993}{809}.} 
 \lref\EK{T. Eguchi and H. Kanno,
 ``Toda lattice hierarchy and the topological
description of the $c=1$ string theory'',
 hep-th/9404056.}
 \lref\DMP{R.Dijkgraaf, G.Moore and M.R.Plesser, 
\np{394}{1993}{356}.
.}
 \lref\GMukhi{D.~Ghoshal and  S.~Mukhi, ``Topological Landau-Ginsburg
model of two-dimensional string theory,'' 
\hepth{9312189}.}
\lref\MVafa{ S.~Mukhi and C. Vafa, ``Two-dimensional Black Hole as a Topological Coset
Model of c = 1 String Theory", \hepth{9301083}.}
\lref\TakTak{K.~Takasaki, T.~Takebe,
``Quasi-classical limit of Toda hierarchy and W-infinity symmetries,''
\hepth{9301070}, \lmp{28}{93}{165}.}
\lref\Krichever{I. Krichever, {\it Func. Anal. i ego pril.},
{\bf 22:3} (1988) 37  (English translation:
{\it Funct. Anal. Appl.} {\bf 22} (1989) 200);
``The  $\tau$-function of the
universal Witham hierarchy, matrix models and topological field 
theories'', {\it Comm. Pure Appl. Math.} {\bf 47} (1992), 
\hepth{9205110}.}
\lref\sixv{I. K. Kostov,  ``Exact solution of the six-vertex model on 
a random lattice'', \np{}{2000}{513}, \hepth{9911023}.}
\lref\kkvwz{ I. Kostov, I. Krichever, M. Mineev-Veinstein,
P. Wiegmann, and  A. Zabrodin, ``$\tau$-function 
for analytic curves", \hepth{0005259}.}
\lref\orlshu{A. Orlov and E. Shulman, \lmp{12}{1986}{171}.}
\lref\Tachtajan{ L. Tachtajan,
``Free bosons and tau-functions for compact Riemann surfaces and 
closed Jordan curves'', {\tt math. QA}/0102164}
\lref\Zabrodin{ A. Zabrodin, ``Dispersionless limit of Hirota 
equations in some problems of complex analysis'',
{\tt math.CV/0104169}.}
\lref\bmrwz{ A. Boyarsky, A. Marshakov,  O. Ruchhayskiy, 
P. Wiegmann, and  A. Zabrodin, ``On associativity equations in 
dispersionless integrable hierarchies", \hepth{0105260}.}
\lref\wz{P. Wiegmann and  A. Zabrodin, ``Conformal maps and
dispersionless integrable hierarchies", \cmp{213}{2000}{523}, 
\hepth{9909147}.}
\lref\Kutasov{D. Kutasov, ``Irreversibility of the renormalization 
group flow in two-dimensional quantum gravity", \hepth{9207064}.}
\lref\FY{ M. Fukuma and S. Yahikozawa,
``Comments on D-Instantons in $c<1$ Strings'',
\pl{460}{1999}{71}, \hepth{9902169}.}
\lref\Nakatsu{
T. Nakatsu, ``On the string equation at $c=1$'', \mpl{A9}{1994}{3313},
\hepth{9407096}.
}
\lref\TakSE{K. Takasaki,
`` Toda lattice hierarchy and generalized string equations'',
\cmp{181}{1996}{131}.
}
\lref\zabmv{M. Mineev-Weinstein and A. Zabrodin,
``Whitham-Toda hierarchy in the Laplacian growth problem'',
\hepth{9912012}.}

\Title{\vbox{\baselineskip12pt\hbox
{SPhT-t01/075}\hbox{ }}}
{\vbox{\centerline
 {String Equation  for }
\centerline{String Theory on a Circle}
 \vskip2pt
}}
 %
\centerline{Ivan K. Kostov\footnote{$^\circ$}{{\tt 
kostov@spht.saclay.cea.fr}}}
\centerline{{\it  Service de Physique 
Th{\'e}orique,
 CNRS -- URA 2306, }}
\centerline{{\it C.E.A. - Saclay,   
  F-91191 Gif-Sur-Yvette, France}}
%
 
\vskip 1cm
\baselineskip8pt{

\baselineskip14pt{
\noindent
We  derive a constraint (string equation) which 
together with the Toda Lattice hierarchy
determines completely   the
integrable structure of the compactified 2D string theory.
 The  form of the constraint  depends on a continuous parameter,
 the   compactification 
 radius $R$.  
   We   show how to use the   string equation  
   to calculate the free energy and the correlation functions in 
   the dispersionless limit. 
   We sketch  the  phase diagram and the flow  structure and point out that there 
are two UV critical points, one of which (the sine-Liouville  string theory) describes infinitely strong vortex or tachyon perturbation.
  }

\Date{July, 2001}

\baselineskip=16pt plus 12pt minus 2pt

\vskip 20pt\noindent{\it $\circ$    Introduction}
 
\smallskip\smallskip

 Bosonic  string theory with a two-dimensional target space,
or $c=1$ string theory (see, {\it e.g.}, the review  \KLEBANOV ), is commonly
  considered as completely solved. In fact,  this is 
  not quite true, and some of the 
   most intriguing questions here are still waiting to be answered.
   One of these questions is whether   strong
   perturbations  of the world sheet action can change the 
   geometry of the target space. According to the FZZ conjecture \FZZ
   ,  the compactified $c=1$ string theory in  
   presence of a sufficiently strong vortex or tachyon  source
   is related by S-duality to a 2d string theory in a Euclidean black
   hole background \refs{\WITTENBH, \DVV}.
   The string theory in a strong vortex  background forms  a new phase,  
   which we call the sine-Liouville phase, because 
   the role of the  Liouville potential  here is  
   played by a sine-Liouville potential\foot{The relation between the topological version of the 2d Euclidean black hole and the c=1 string compactified at  the self-dual radius
has been previously pointed out by Mukhi and Vafa \MVafa .}.

  A vortex perturbation of the 2d string theory 
  is equivalent to a time-dependent tachyon perturbation 
  in the T-dual theory. The simplest nontrivial example of such a 
  perturbation is the sine-Gordon model coupled to quantum gravity,
  which was first   studied by G. Moore \refs{\MOORE}. 
  Some interesting results followed for the general case,
  in particular  the discovery of the 
   integrable structure of the Toda Lattice hierarchy  \DMP.

 Recently, this problem has been reconsidered in \KKK ,  where
 the fact  that  the flows generated by allowing vortices 
 on the world sheet  are  described by  the Toda Lattice hierarchy was used.
 In this paper we complete the approach of \KKK\ by deriving a 
 constraint (a "string equation"), which, together with the 
 Toda Lattice hierarchy, completely determines the theory.
 We consider in more detail the dispersionless (genus zero) limit and
 show how to use  the string equation to perform calculations.
  We then check  that the string equation correctly reproduces the    free energy
in the presence of a pair of vortex  operators obtained in  \KKK, as well as the
   one- and two-point correlators in the Liouville phase, recently calculated in  \AK.
   Finally,  we discuss the  phase
  structure  of the theory and 
   give a qualitative description of the  sine-Liouville phase.
   
\noindent{\it $\circ$    Tachyon and vortex sources in the compactified 
 $c=1$ string 
 theory}
\smallskip 

  Euclidean 2D   string theory in a flat background 
 is described on the string worldsheet by the
 conformal field theory of
 a massless scalar $x$ coupled to a $c=25$ Liouville field $\phi$, 
 with worldsheet Lagrangian
  \eqn\conepert{
 \CL={1\over4\pi}[(\partial x)^2 +(\partial\phi)^2
 -2\hat R\phi+ \mu  e^{-2\phi}] + \CL_{\rm ghost}.} 
 If the $c=1$ coordinate is  compactified
 at radius\foot{We will use units in which the T-duality acts as 
 $R\to 1/R$ and thus  $R_{\rm  KT }=2$.}
 $R$,  then the theory has a discrete spectrum and
the excitations carrying the
 momentum    and the winding modes are represented by the 
   tachyon operators $\CT_n$   and the vortex operators $\CV_n$
($n=0, \pm1, \pm 2, ...$)
    \eqn\tachb{ 
   \CT_{n}    \sim 
   \int _{\rm world \ sheet } e^{ inx/R}  e^{(|n|/R-2)\phi}}
  \eqn\verb{  
   \CV_{n}  \sim 
   \int _{\rm world \ sheet }e^{in\tilde x R}   e^{(|n|R-2)\phi },}
 where the field $\tilde x$ is T-dual to $x$.
    The vortex operators can be created as  topological defects  
 describing localized winding modes: a vortex of charge $nR$ 
 is located at the endpoint of a line 
along which the time coordinate has discontinuity $2\pi  nR$. 

\vskip 20pt\noindent{\it $\circ$    Matrix model formulation and Toda Lattice symmetry }
  \smallskip\smallskip
   
The $c=1$ string theory compactified at length $2\pi R$
can   be constructed as a  large $N$ matrix model
(Matrix Quantum Mechanics),
which can be viewed as a dimensional reduction of an 
 $N$-color  2d Yang-Mills theory
(for details see  the review  \KLEBANOV).   After being dimensionally reduced to the circle  
$x+2\pi R \equiv x$, the YM theory is described in terms of two hermitian
 $N\times N$   matrix  fields,
the  Higgs field  $M=M_i^j(x)$ and the gauge field $A=
   A_i^j(x)$.
It is 
more convenient to consider the grand
canonical partition function, in which the chemical potential $\mu$
plays the role of a cosmological constant:
\eqn\prtf{
\CZ ( \mu,  R, \hb) = \sum_{N=0}^{\infty}e^{-\bh 2\pi R\mu N}
\int_{^{A(x+2\pi R)=A(x)}_{M(x+2\pi R)=M(x)
}} \CD M\CD A\  
e^{ -{1\over \hbar} \Tr\  
\int ^{ 2\pi R} _0   (-\hf [i\p_x + A,\  M]^2 
  +\hf M^{2})dx}.}
We have introduced explicitly the Planck constant $\hb$, which 
is also the  string interaction constant. 
The dominant contribution of the sum over $N$ comes from
$N\sim 1/\hb$.
In order to make sense of the path integral,
the   measure $\CD M$ should be stabilized by  an appropriate  cutoff,
for example an infinite  potential wall placed far from the 
origin.
 
The tachyon  operators are represented as the discrete 
momentum modes of the    Higgs field $M(x)$\foot{This expression for the 
tachyon operators in the scaling limit is due to V. Kazakov.}
\eqn\tachn{
\CT_{n}\sim \int _{0}^{2\pi R} dx \ e^{i{n\over R} x} \left[M(x)\right]^{|n|\over R }
}
while the  vortex operators can be
 constructed as the moments of the   holonomy factor of the
 gauge field $A(x)$ around the spacetime circle \KKK\
\eqn\vortx{
\CV_{n}\sim \Tr\ \Omega^{n}, \quad \Omega=
\hat T \exp \int _{0}^{2\pi R}dx A(x) .
}
Note that, in spite of the asymmetry in the definition of the tachyon and vortex 
operators, their expectation values and correlation functions are 
related by
the T-duality  ($R\to 1/R$)\foot{The T-duality can be made explicit 
if the matrix model on a circle \prtf\ is  considered as the
 limit of Yang-Mills theory defined on a two-dimensional torus, when 
one of the periods tends  to zero. 
Then the  tachyon and 
vortex operators \tachn-\vortx\  are constructed as   Polyakov  loops 
winding $n$ times around   one of the two periods of the torus.}.

 The integral over 
the gauge field
$A$ can be done explicitly and the resulting integral with respect to
the eigenvalues of the Higgs field $M$ 
can be written as the partition function of  free nonrelativistic
fermions defined on the spacetime circle.  
This remains true  also if  a  purely tachyonic source
$\sum_{n}\l_{n}\CT_{n}$ 
is added to the exponent in \prtf .
Using the properties of the amplitudes for  tachyon  scattering
  \MPR ,
 Dijkgraaf, Moore and Plesser showed in \DMP\  that 
the partition function 
\prtf\  with purely tachyonic source added 
is a $\tau$-function of 
the Toda Lattice   hierarchy \UT .         
 
  On the other hand, 
  if the theory is  perturbed by a   purely vortex source
 $ \sum_{n\ne 0} t_{n} \CV_n    $, then the 
 integral with respect to the Higgs field 
 $M$ can be done exactly\foot{This observation was first made by
  by Boulatov and Kazakov
 \BULKA, who used it  to study 
the non-singlet states of the $c=1$ string theory.} and 
one obtains an effective
 one-plaquette gauge theory whose only variable is the 
unitary matrix $\Omega$, the  holonomy factor around the 
spacetime circle \KKK\foot{This matrix model
 belongs to  a class of integrable 
  models \refs{\NKK , \HKK, \sixv}
  representing various generalizations of  the Gaudin gas \GGG.}. 
  The partition function of this theory can be expressed 
  in terms of free fermions defined on the
  unit circle and is   shown to be again a
  $\tau$-function of the Toda Lattice
  hierarchy.

 In this way the $c=1$ theory perturbed by a purely vortex or purely
 tachyonic source is completely integrable and its  integrable 
 structure is described by   the Toda Lattice hierarchy\foot{The 
 integrability 
is  however lost if the theory is 
 perturbed by {\it both} tachyon and vortex sources.}.  
The Toda Lattice hierarchy  is formulated in 
 terms  of the  ``time'' variables $t_{\pm 1}, t_{\pm 2}, ...$
and   a discrete
   ``spatial'' variable 
 $s\in \hb \Z$.  
It is natural to split the time variables into two sets
 $t  = \{t_{ n} \}_{n=1}^{\infty}$ and  
  $\bar t  = \{\bar t_{ n}\}_{n=1}^{\infty}$  where $\bar t_n\equiv 
t_{-n}$.
The correspondence with  2D string theory 
 requires to consider 
 a {\it complex}    variable $s$, which is related to the 
 cosmological constant $\mu$ as
 $s = \hf - i \mu $. The Planck constant $\hb$, which 
 appears as the spacing unit of the difference operators in the Lax 
 formalism, plays the role of string interaction constant.
 The exact statement, made in \DMP\ on the basis of the analysis of 
 the tachyon S-matrix in \MPR, is\foot{The derivation of \DMP\ was 
  critically  reconsidered in \MukhiImbimbo.}
 that the  free energy of the 
 string theory
 \eqn\frnE{\CF (\mu, t, \bar t)= \sum _{g\ge 0} \hbar ^{2g}
 \llangl  e^{\sum_{n_{\ne 0} }  t_n \CV_n }    \rrangl_{ g}}
 where $\llangl\rrangl_{g}$ means the  connected expectation value 
 on a genus-$g$ worldsheet, is related to the $\tau$-function of the 
 Toda Lattice hierarchy as
 \eqn\tauffr{ \tau = e^{\CF /\hb^{2}} .}
  Thus the free energy of the string theory satisfies an
   infinite chain of PDE, the first of which is the Toda equation
  \eqn\todaeq{
  {\p\over \p t_{1}}{\p\over \p  \bar t_{1}} \CF(\mu)+
 \exp\left({1\over \hb^{2}}[\CF(\mu+i\hb)+\CF(\mu-i\hb)
  -2 \CF(\mu)]\right)=0.}
 This equation determines the flow  in the compact $c=1$ string theory 
perturbed   by the   lowest vortex operators $\CV_{\pm 1} $.
In \KKK , the free energy on the sphere and on the torus was obtained as the solution of
   \todaeq\ satisfying at $t=\bar t=0$  the boundary condition
\eqn\boundc{
\CF(\mu)= -\hf R \mu^2\log\mu - \hb^2 {R+R^{-1}\over 24} \log\mu +\CO(\hb^2).}
The   string susceptibility 
    $\chi  = \p_\mu ^2\CF$ at genus zero  was obtained as
     a solution of a simple algebraic equation, 
     \eqn\susca{\mu e^{\chi /R}+ (R-1)t_{1}\bar t_{1}
      e^{(2-R)\chi /R}=1,}
     which resums  
     the  perturbative expansion found in   \MOORE .  
  The same approach was subsequently applied    to calculate 
      the vacuum expectation values of the vortex operators and the 
  two-point correlators \AK. 

The algebraic equation \susca\  suggests strongly that
   an operator  constraint (a ``string equation'')  similar to those found for  
    the $c<1$ matrix models \MikeKDV\   might exist.
  Below  we derive  the string equation, which leads to
  \susca, but before doing that we will 
  briefly recall the Lax formulation 
  of the Toda hierarchy \UT.

\vskip 20pt\noindent{\it $\circ$     Lax formalism}
\smallskip\smallskip
       
  The  Lax formalism is based on finite-difference operators 
where the Planck constant $\hb$ (which in our interpretation plays 
the role of string interaction constant)
emerges as a spacing unit \TakTak . 
The Lax operators $ L $ and
$\bar L$
are  defined as series expansions in the shift operator
\eqn\omdf{ \oo =
  e^{i\hb \p/\p \mu}}
with coefficients depending on  $\mu $ and 
  the couplings $t$ and $\bar t  $
\eqn\Lopl{
 L =r  \oo +
\sum_{k=0}^{\infty}u_k \oo^{-k},\ \ \ 
 \bar L = \oo^{-1} \bar r +
\sum_{k=0}^{\infty}
 \oo^{k} \bar u_{k}.
}
The commuting flows along the ``times"  $t_n$  and $\bar t_n$
are generated by the operators 
\eqn\hashka{
 H_{ n} =  (L^n   ) _{> 0} +\hf (L^n   ) _{0} 
 , \ \ \ \bar H_{ n} =  (\bar L^n   ) _{< 0}+\hf  (\bar L^n   ) _{0}
 , \ \ \ 
(n=1,2,\ldots)
}
where the symbol $   ( \ \  )_{^{>}_{<}}$ means the positive
(negative) parts of the series in the shift operator
$ \oo $ and $   ( \ \   )_{ 0}$ means the
constant part\foot{ We  use slightly different notations than in 
\UT\  because we would like to
preserve the symmetry 
between $  L$ and $    \bar L$, which    is appropriate for the 
double 
scaling limit of the matrix model. 
}. 
 If we define the ``covariant derivatives''
 \eqn\covdi{ \nabla_n = \hbar {\p \over \p t_{n} } - H_{n} \ \ \  
  (n = \pm 1, \pm 2, \ldots),}
then the   Lax  equations 
  \eqn\SSE{ [\nabla_n ,  L]  =  [\nabla_n  ,\bar L] =0  ,}
are equivalent to the zero-curvature
  conditions        
\eqn\zerocur{ [\nabla_m,\nabla_n] =0 .
 } 
  %
  %
  The Lax operators  $ L, \bar L$  can  be thought as the result of 
  dressing  transformations
   \eqn\Osoev{ L=    W  \ \oo 
   \  W^{-1} ,\ \   \bar L=  \bar   W  \ \oo ^{-1}
   \ \bar W^{-1} 
   ,}
    where the dressing operators $ W$ and $ \bar W $  have  
    series expansions
   \eqn\drexp{
 W =  
 \sum_{n\ge 0} w_{n}    \oo^{  - n} ,\ \ 
  \bar W =   
 \sum_{n\ge 0}\bar  w_{n}    \oo^{   n}
 .}
 Lax-equations can be converted into Sato equations 
 for the dressing operators
 \eqn\saioeq{
 \nabla _n W = \nabla_n \bar W = 0, \ \ n=\pm 1, \pm 2, ...}

 The dressing operators are related to the $\tau$-function
  \tauffr\  by
 \eqn\dropt{\eqalign{
  z^{-i\mu} W z^{i\mu}&=\tau^{-1}  
   e^{-\hb D_{t}(z) - \hf i \hb \p_{\mu}}  \tau\cr
   \bar z^{-i\mu}\bar W \bar z^{i\mu}&=\tau^{-1} 
   e^{-\hb D_{\bar t } (\bar z) + \hf i\hb \p_{\mu}}  \tau
   }
 }
where we introduced the spectral parameters $z$ and $\bar z$ 
and the symbols
\eqn\DDbar{
D_{t}(z)  = \sum_{n} {1\over n\  z^{n}} \ {\p \over \p t_n }, \
 D_{\bar t }(\bar z) = \sum_{n} {1\over n\ \bar z^{n}}
\ {\p \over \p \bar t_{n} } .  
}
The $\tau$-function contains all the  data of the Toda system.
  For example, the first coefficients  in
the expansion of  Lax operators \Lopl\ and the dressing operators
  are 
related to the $\tau$-function as
\eqn\rrw{r\bar r (\mu) 
=  { \tau(\mu+i\hb)\tau(\mu-i\hb) \over
\tau^{2}(\mu) } .
}
 
The dressing operators are determined by \Osoev\ up to a factor that commutes with
the shift operator. In particular,  eq. \Osoev\ is satisfied by  the   
 wave operators
\eqn\CWWW{
\CW = W \exp\left({1\over \hb} \sum_{n\ge 1} t_n \oo^n\right),
\ \ \ \ 
\bar \CW = \bar W \exp\left( {1\over \hb} \sum_{n\ge 1} \bar t_n 
\oo^{-n}\right)}
which have the following important  property.
The operator
  \eqn\wwbar{ \CA = \CW^{-1}\bar \CW}
  does not depend on $t$ and $\bar t$ 
  (see Theorem 1.5 in ref. \UT\ and Proposition 1.2 in ref.\Takebe).  
The operator  \wwbar\    characterizes the particular  solution 
  of the Toda Lattice hierarchy.
 
\vskip 20pt\noindent{\it $\circ$     Orlov-Shulman operators}
\smallskip\smallskip

   The  Lax system can be extended 
    by adding 
Orlov-Shulman  operators 
 \orlshu\  
 \eqn\Osoem{\eqalign{
 M & = W \left(\mu +  i\sum_{n\ge 1}  n\ t_{ n}\
 \oo^{ n} \right)W ^{-1} 
  , \cr
  \bar M & = \bar W \left(\mu  -  i\sum_{n\ge 1}  n\    \bar t_{ n}\
 \oo^{- n} \right)\bar W  ^{-1} .  \cr
 }}
 The Lax and Orlov-Shulman operators are expressed in terms of 
 the wave operators \CWWW\ as
 \eqn\LLMM{\eqalign{
 L&=\CW \oo  \CW^{-1} , \ M= \CW \mu  \CW^{-1}\cr
 \bar L&=\bar  \CW \oo ^{-1} \bar \CW^{-1}  , \ 
 \bar M=\bar  \CW  \mu \bar \CW^{-1} .}
 }
 Therefore  the  pairs $L, M$ and $\bar L, -\bar M$
satisfy the same commutation relations as the pair $(\oo, \mu)$:
\eqn\lmpair{[\oo, \mu]= i\hbar \oo, \ \ \ [L, M]= 
  i\hbar L, \ \ [\bar L, \bar M] = - i\hbar \bar L.
}
 
Since the ``spatial" parameter of the Toda system $s= \hf -i\mu$  
is continuous in our consideration, one can also consider arbitrary powers $\oo^\a$ of the shift 
operator and generalize \lmpair\  to
\eqn\LMpair{ [\oo^\a , \mu]= i\a \hbar \oo^\a,\ \ \ 
 [L^\a, M]= 
  i\a\hbar L^\a, \ \ [\bar L^\a, \bar M] = - i\a\hbar \bar L^\a.
}

\vskip 20pt\noindent{\it $\circ$     Gaussian field representation}
\smallskip\smallskip

The Orlov-Shulman operators  can be represented as the   currents  of 
the gaussian field
$\Phi (z, \bar z)= \Phi(z) +\bar\Phi(\bar z)$ whose left and right 
chiral components are defined by
\eqn\PHI{\eqalign{
 \Phi (z )&=  \mu \log z - \hf i\hb ^2 {\p\over \p\mu}+
 i \sum_{n=1}^\infty z^n  t_n   -i \hb^2 {z^{-n}\over n} {\p\over \p t_{n}}
 \cr
\bar\Phi(\bar z)&= \mu \log\bar z  + \hf i \hb^2 {\p\over \p\mu}
+ i\sum_{n=1}^\infty  \bar z^n  \bar t_n-i\hb^2
{\bar z^{-n}\over n} {\p\over \p \bar t_{n}}.
}} 
 Indeed, introducing the spectral parameters $z $ and $\bar z $,
we can write \dropt\  in the form
\eqn\expP{
\eqalign{
   \CW z^{i\mu}&= \left\langle  e^{  \Phi (z) /\hb}
      \right\rangle\cr
  \bar \CW  \bar z^{i\mu}&= \left\langle  e^{  \bar  
\Phi (\bar z) /\hb}
      \right\rangle
  }}
where  for any differential operator $\CO$ in $t$, $\bar t$  and $\mu$
we denote   $\langle \CO \rangle = \tau^{-1}  \CO
      \tau$.
From \expP\ and    \LLMM \ we find
\eqn\loopoprs{\eqalign{
  z^{-i\mu} M z^{i\mu}&= \left\langle z \p_z   \Phi (z) 
      \right\rangle\cr
   \bar z^{-i\mu}\bar  M \bar z^{i\mu}&= \left\langle  \bar z 
\p_{\bar z}
     \bar \Phi (\bar z)   \right\rangle  . }
 }
Eq. \loopoprs\  yields, together with  \hashka, 
the following expansions for the Orlov-Shulman operators
as Laurent series in  $L$ and $\bar L$ 
\eqn\MMexp{\eqalign{
M  &= \mu  + i\sum _{n=1}^{\infty} 
\left( n t_{n} L ^{n}+
   v_{  n}   L ^{-n}\right)\cr 
 \bar   M & = \mu -i\sum _{n=1}^{\infty}
\left( n \bar t_{n} \bar L ^{n}+
   \bar v_{ n}  \bar L ^{-n}\right)
   }
   }
 with  coefficients $v_{n}$  and $\bar v_{n}$  equal to the 
expectation values of the vortex operators
\eqn\vevv{ v_{n} = {\p\CF \over \p t_{n}} =
\langle \CV_{n}\rangle ,\ \ \ \ \bar v_{n} = {\p\CF\over \p \bar 
t_{n}}
 =\langle \bar \CV_{-n}\rangle .}
The operators $\Phi(z)$ and $\bar\Phi(\bar z)$ can interpreted as
creating and annihilating  world-sheet boundaries  and 
  the spectral parameters $z$ and $\bar z$ play the role of 
boundary cosmological constants.
Similarly,  the operators $M$ and $\bar M$   create
and    annihilate   boundaries  
with marked points  because of the derivatives in $z$ and $\bar z$.
In this way 
 Orlov-Shulman  operators allow   the spectrum of local scaling  operators on the world sheet to be studied. 

 For sufficiently  weak  perturbations, these are the vortex 
and anti-vortex operators associated with the expansion 
\MMexp\ in integer powers of $z=L$ and $\bar z = \bar L$.
 For strong perturbations, 
where the series \MMexp\  do not converge, the operators
$M$ and $\bar M$ can be given meaning  {\it via} analytic 
continuation. We will return to this 
point at the end of the paper.

\vskip 20pt\noindent{\it $\circ$    The ``string   equation"}
   \smallskip\smallskip

 Now we are ready to proceed with the derivation of the constraint
which, together  with the Toda Lattice hierarchy,   defines
 the integrable structure of the compactified string theory.
Such  constraints are   commonly called
  ``string equations''.

  Let us first consider the case  of the unperturbed
  theory ($t =\bar t =0$). In this case the all-genus free energy of the  2d  string theory compactified at radius $R$ is given by the integral
  \KLEBANOV 
  \eqn\frenergy{\CF(\mu)=\hb^2
\log\tau(\mu)=
{\hb^2 \over 4 }\
\Im \int_{-\infty}^\infty {dy\over y} \ {e^{-{2i R}  y\mu/\hb}\over 
\sinh ( yR)
\  \sinh ( y)}}
and therefore satisfies 
the functional 
equation  
\eqn\diffeq{ 4
\sin \left(\hb  {\p_\mu /2 R}\right)
\ 
\sin \left(\hb  {\p_\mu /2 }\right)\ \log\tau = \log(1/ \mu).
}
  Now we would like to rewrite \diffeq\ as 
  an algebraic relation  for  the   Lax operators.
  In the case of zero potential, only the first term in the expansions
  \Lopl\ survives
  \eqn\llboo{ L  =  r \oo = 
    W \oo  W^{-1}
  , \ \ \bar L =
  \oo^{-1}  \bar r  
  = \bar  W \oo ^{-1} \bar W^{-1}  
  }
  where $W(\mu) $ and $\bar W(\mu) $ are ordinary functions.
  By   \dropt\  we  find for their ratio 
  \eqn\drfun{ 
    { W(\mu)\over \bar W(\mu)}= {\tau(\mu - 
  i\hb/2 )\over \tau(\mu +  i\hb/2 )},}
      and  therefore the functional equation \diffeq\  
   is equivalent to
   \eqn\drf{  {  \bar W(\mu + i\hb/2R)W(\mu - i\hb/2R)
   \over W(\mu + i\hb/2R) \bar W(\mu - i\hb/2R)}=\mu,}
    which means that    the operator $ A= W^{-1}\bar W $ 
     satisfies the following   
     identities 
\eqn\operSE{\eqalign{
A\oo^{-1/R} A^{-1}\oo^{1/R} &=  \mu  - i {\hb/ 2R} , \cr
 A^{-1} \oo^{1/R} A\oo^{-1/R} &= \mu + i {\hb/2 R}   .\cr} }
 A third identity follows from the fact that for $t=\bar t=0$ 
 the  dressing operators do not contain shifts and therefore commute with $\mu$:   \eqn\meqmbar{
  A\mu A^{-1}=\mu.}
Further, at  $t =\bar t =0$  the dressing operators $W$ and $\bar W$ 
coincide with the 
wave operators $\CW$ and $\bar \CW$ defined by \CWWW\ and we can replace
the operator $A$ in  \operSE -\meqmbar\ by $\CA= \CW^{-1}\bar \CW$.
Therefore   the identities \operSE-\meqmbar\ 
     actually hold for {arbitrary}  couplings $t$ and $\bar t$. 
    
   Eqs. \operSE\ and \meqmbar\ can be formulated as 
    algebraic relations between      $L, \bar L$, $M$ and $\bar M$, namely
     \eqn\LMSE{ \eqalign{
\bar L^{1/R} L^{1/R}= \bar M -{i\hb\over 2R},\cr
L^{1/R}\bar L^{1/R}=  M +{i\hb\over 2R},}
}
    \eqn\meqmbar{
 M=\bar M.} 
 Thus we arrive at the constraint, which allows to express 
 the Orlov-Shulman operators as bilinears of Lax operators
 \eqn\MMMM{M=\bar M = \hf\left(L^{1/R}\bar L^{1/R}+ \bar 
  L^{1/R}L^{1/R}\right).}
 Given the ``string equation" \MMMM,  the canonical commutation relations
\LMpair\ are equivalent to the following   constraint   for the  Lax operators
\eqn\seL{[L^\a, \bar L^{1/R}]=i\a\hbar L^{\a-1/R}, \ \ \ 
[\bar L^\a, L^{1/R}]=- i\a\hbar \bar  L^{\a-1/R}.}


 \vskip 20pt\noindent{\it $\circ$    The dispersionless   limit $\hb\to 0$ }
    \smallskip\smallskip
	   
In the sequel we will concentrate on the genus zero case and 
consider the
 dispersionless 
  limit $\hb \to 0$ of Toda hierarchy, which is a   particular case 
   of the universal Whitham hierarchy introduced by Krichever
   \Krichever . The solutions of the dispersionless  hierarchy 
   can be parametrized by canonical transformations in a 2D phase 
   space so that any solution is determined by the choice of a 
   canonical pair of variables \TakTak .

   In the   limit  $\hb\to 0$, the operator $\oo$
 can be considered
as a classical variable $ \o $, conjugate to $\mu$  
\eqn\ccrom{ \{  \o  , \mu\} =   \o }
where the Poisson bracket is defined by 
\eqn\poissb{ \{ f, g\} = i\o( \p_{\o}f
 \p_{\mu}g - \p_{\o}g \p_{\mu}f),}
and the Lax operators $L,\bar L$ are considered as   c-functions
of the  phase space coordinates  $\o$ and $\mu$.  The Hamiltonians 
as functions of $\o$ and $\mu$ give the expectation values of the 
vortex operators, $H_n=\p_{t_{n}}\CF =\langle \CV_n\rangle$.

 The Lax equations are 
equivalent to the exterior differential relations
\eqn\extlx{d\log L \wedge dM = d\log \bar L \wedge d\bar M
= d\log\o \wedge d \m +i\sum_{n\ne 0} d H_n \wedge d t_n
} 
which   imply  the existence of    functions 
$S(L, \m, t,\bar t)$ and $\bar S (\bar L, \m, t,\bar t) $ 
whose differentials are given by
\eqn\zVKB{ \eqalign{
d S &=
    M  d\log L  
+
\log  \o \  d\mu
  +i\sum_{n\ne 0}  H_n   dt_n
 \cr  d\bar S&=
   \bar M d\log   \bar L  
+
\log  \o \  d\mu
  +i\sum_{n\ne 0}  H_{n}  dt_n.
}
}
The functions $S$ and $\bar S$ can be thought of as the generating 
functions 
for the  canonical transformations
$(\o,\mu)\to (L, M)$ and $(\o,\mu)\to (\bar L, \bar M)$. The new 
coordinates $z=L( \o, \m)$  and
 $\bar z = \bar L(\o, \m)$ are analytic functions of $\o$, 
 given by the $\hb\to 0$ limit of the
  expansions  \Lopl.  The 
classical Hamiltonians  $H_n(\o,\mu)$ associated with the ``times''
$t_n$ are   related 
$z=L(\o)$ and  $\bar z =\bar L(\o)$  by \hashka  .  

 Conversely, the coordinate $\o$ can be considered as a 
 meromorphic function of either $z$ or $\bar z$
\eqn\Omzzb{ \o = e^{\p_{\mu}S(z)} = e^{\p_{\mu}\bar S(\bar z)}.}
Eq. \Omzzb\ defines the canonical transformation between $(L, M)$ and 
$(\bar L, \bar M)$.

   Further, the functions $S(z)$ and $ S(\bar z) $  are 
equal to the expectation values of the chiral components
of the gaussian field 
\PHI 
 \eqn\ssbarz{ \eqalign{S(z)  & =\langle \Phi(z)\rangle
\cr
    \bar S(\bar z)  &= \langle\bar  \Phi(\bar z)\rangle
 }}
and the functions representing  
Orlov-Shulman operators are given by the derivatives
\eqn\MMSS{M(z) = z\p_{z}S(z), \  \bar M(\bar z) = \bar z\p_{\bar z} 
\bar S(\bar z).}

In the dispersionless  limit  we can write     \MMMM\    and \seL\ as
\eqn\cexSE{M = \bar M=  L^{1/R} \bar L^{1/R}
  }
  and
\eqn\cSEQ{
  \{ L^{1/R} , \bar L^{1/R} \}= 
  {1/R} .}
 The second constraint is a consequence of the first one and the 
 canonical commutation relations 
\eqn\ccrLL{\{L,M\}=L, \ \ \{\bar L, \bar M\} = - \bar L.}
   
 The dispersionless  ``string equation"  \cexSE\
  generalizes the constraint    found for the self-dual 
radius  ($R=1$) by Eguchi 
  and Kanno \EK, and for all integer values of $
 \beta = 1/R$ by Nakatsu \Nakatsu . 
In this last case one can rewrite the ``string equation" as a
linear constraint ($W_{1+\infty}$ constraint) 
on the $\tau$-function \TakSE.
 
 \vskip 20pt\noindent{\it $\circ$    Two-point correlators in the dispersionless limit}
 \smallskip\smallskip

 Knowing the  functions $z(\o)$ and $ \bar z(\o)$ 
 or their inverse functions $\o(z)$ and $\o(\bar z)$,  
 we can evaluate, 
 using  \Omzzb \ and \ssbarz,  the 
  two-point functions $ \p_{\mu}\p_n\CF$:
 \eqn\muen{\p_\mu\p_n \CF =  {1\over 2\pi i} \oint_{\infty} 
 z^{n} d\o(z),
 \ \  \p_\mu\p_{-n} \CF =  {1\over 2\pi i} \oint _{0} 
 \bar z^{n} d\o(\bar z)\ \ \ \ (n\ge 1).}
   The  generating functions for the
two-point correlators $\p_{m}\p_{n}\CF$ and $\p_{m}\p_{-n}\CF$ $(m, n>0)$
are given by
\eqn\tpca{\eqalign{\hf \p_{\mu}^{2}\CF+
D_{t}(z_{1}) D_{t}(z_{2}) \CF
&= 
\log{\o(z_{1})-\o(z_{2})\over z_{1}-z_{2}}\cr
}
}
 \eqn\tpcb{\eqalign{ D_{t}(z_{1})  D_{\bar t}(\bar z_{2}) \CF&=
- \log\left( 1- { 1  \over \o(z_{1})\o (\bar z_{2}) }
\right) \cr 
}
}
where the operators $  D_{t}(z) $ and  $ D_{\bar t}(\bar z)$ are defined by 
\DDbar\ and the functions $\o(z)$ and $\o(\bar z)$ are defined by \Omzzb.
Eqs. \tpca\ and \tpcb\ are obtained directly from the 
dispersionless 
Hirota equations for  Toda hierarchy \TakTak\ and hold independently 
of the constraint \cSEQ.
  
 \vskip 20pt\noindent{\it $\circ$    Landau-Ginsburg description }

The topological description of the $c=1$ string theory proposed for 
the
self-dual radius in  \GMukhi\ and \HMP\   can be readily generalized for any 
value of the compactification radius. It was shown in 
\bmrwz\ that the WDVV equations  are  a consequence of  the 
dispersionless Hirota equations.

 \vskip 20pt\noindent{\it $\circ$    Relation to conformal maps}
     \smallskip\smallskip

   The  solutions of the dispersionless Toda hierarchy
can be given a geometrical interpretation as 
conformal maps from  simply connected domains in the complex plane
to the unit disk. This correspondence was  first established 
for a particular solution of the hierarchy \refs{\wz , \kkvwz} and 
then  
for any generic solution in \Zabrodin .

The general form of a conformal map from the exterior   of the unit 
disc   to  a simply connected domain 
$\CD$  containing the point $z=\infty$  is given by the first  
 series expansion in \Lopl.  
The parameter $r$, which  we will assume to be real and positive,
is called the conformal radius of the domain $\CD$.
 Let us further assume that  the couplings $t_n$ and $\bar t_n $ are 
complex conjugate. Then,
 along the boundary of the unit disc $\bar \o=1/\o$, 
 $\bar z = \bar L(\o)$  and   $z=L(\o)$ are complex conjugates  
and define a smooth curve $\g$ in the complex $z$-plane, which is the boundary 
of the domain $ \CD$. 
 If $t=\bar t=0$, then $\g$ is  a circle of radius $r = \mu^{R/2}$;
   in   the general case   
the form of the contour depends on the values of the couplings $t_n$.
The couplings $t_n \ ( n\ne 0)$ and $\mu$  can be thought of as a set 
coordinates in the space of closed curves. 
The coordinates of the curve $\g=\p \CD$  
are given by the moments of the domain $ \CD$   
 with respect to the measure $\s= \s(z,\bar z)dz\wedge d\bar z $  
\eqn\moments{t_n =-{ 1\over \pi n}  \int_{ \CD}  z^{-n} \s(z,\bar 
z)d^2z,
\ \ \bar  t_n =-{ 1\over \pi n}  \int_{ \CD}  \bar z^{-n} \s(z,\bar 
z)d^2z,
\ \ \ 
\mu = { 1\over \pi R}  \int_{ \CD}   \s(z,\bar z)d^2z
}
and the free energy is
$$\CF = {1\over  2\pi ^2} \int _{ \CD} d^2z \int _{ \CD} 
d^2\z\ \ \s(z,\bar z)\
  \log| z^{-1}-\z^{-1}|^2\ 
\s(\z,\bar \z) 
.$$
 In \wz\  it is assumed that the density is homogeneous, which 
in our case is so only at the  selfdual point $R=1$.
The case of generic density  $\s(z,\bar z)$ was worked out in 
\Zabrodin.

The density function is fixed by the constraint imposed on the Toda 
system.
In our case,  the string equation \cSEQ\  means that if we 
parametrize the 
phase space by $z=L$ and $\bar z = \bar L$, then
the volume form  is 
$ \s(z,\bar z)dz\wedge d\bar z = R d(z^{1/R})\wedge d(\bar z^{1/R})$.
This means that the density function is 
\eqn\densitys{\s(z,\bar z)= {1\over R}\  (z\bar z)^{1-R\over R}.  }
 If we introduce the potential
 $U(z,\bar z)$ such that
$ \s(z,\bar z)= \p\bar \p U(z,\bar z)$, the   couplings 
$t$, $\bar t$ and the cosmological constant $\mu$ 
are
given by contour integrals along the boundary $\g = \p \CD$   
\eqn\contm{t_n ={ 1\over 2\pi i n }  \oint_{\g}  z^{-n}  \p_z U  dz,
\ \ \bar  t_n ={ 1\over 2\pi i n }  \oint_{\g}  \bar z^{-n}  \p_{\bar 
z} U  d\bar z,\ 
 \mu = {1\over 2\pi i R}  \oint_{\g}    \p_z U  dz.}
In our case 
$U(z,\bar z)=R\  (z\bar z )^{1/R}.$
The derivative of the potential can be considered as a
 meromorphic function in
the vicinity of the curve $\g$  and is related to the Orlov-Shulman 
operator by
$z\p_z U(z,\bar z) = M(z)$.
Similarly,  $\bar z\p_{\bar z}  U(z,\bar z) = M(\bar z)$.
The dispersionless string equation \cSEQ\  is equivalent to
$\{L, \bar L\} = R (L\bar L)^{1-R\over R}= \s^{-1}(L, \bar L)$.

The  formulation of the dispersionless limit of the theory  as a 
conformal map  problem is  analogous to the now standard geometrical 
description of the tachyon excitations  in the 2D string theory  
within   MQM.  The tachyons there appear as
small fluctuations of the profile of the Fermi surface \POLCHINSKI.
In our case, similarly, a small change of the couplings $t_n$ 
 leads to  
a small deformation of  the boundary of the domain $\CD$.  Typically 
the deformations of the
domain satisfy a $W_{\infty}$ symmetry, but a local realization of
this symmetry  exists only for $R=1$.  In order to have the standard 
realization of the $W_{\infty}$ symmetry, we have to go to  coordinates
 $y=z^{1\over R}, \bar y = \bar z^{1\over R}$.  In these  coordinates
 the density is homogenuous, but the conformal map has a conical
 singularity in the origin.   
 
    \vskip 20pt\noindent{\it $\circ$    How to use the string equation} 
    \smallskip\smallskip

It is convenient to rewrite the expansions \Lopl\  as
\eqn\LLbar{\eqalign{
L^{1 / R}&=  (r \o)^{1\over R} 
\left( 1+a_{1} / \o + 
a_{2}/\o^{2}+\ldots \right)  ,\cr 
 \bar L^{1 / R} &= (\bar r / \o)^{1\over R} 
 \left( 1+ \bar a_{1}\o+ 
 \bar a_{2}\o^{2 } +\ldots 
\right),  }
 }
where $a_k, \bar a_k$ are some functions of the couplings
$t_k, \bar t_k$ $(k=1,2,...)$ and $\mu$.
 The constraint \cexSE\  then gives
\eqn\strq{  (r \bar r)^{1\over R}  \left( 1+a_{1}  \o^{-1}+ 
a_{2}\o^{-2}+\ldots \right)
  \left( 1+\bar a_{1}\o + \bar a_{2} \o^{2 } +\ldots 
\right)   = M(\o).}

In order to use \strq\ we need to   express the coefficients of the
Laurent series $M(\o)$
in terms of the couplings $t_{\pm1}, ..., t_{\pm n}$ and $\mu$.
This  can be done using the expansions \MMexp\
as follows.
It is easy to see that if we consider a perturbation by
$t_{\pm1}, ..., t_{\pm n}$, then all the coefficients but 
$a_1,...,a_n$ and
$\bar a_1, ... \bar a_n$ in \LLbar\ are zero.
Then the  coefficients of the positive [negative]  powers of $\o$ are 
obtained by
plugging $M(\o) = M(L(\o))$ $[M(\o) = \bar M(\bar L(\o))$] into \MMexp.
 Comparing the coefficients  of  $\o^{\pm 1}, ..\o^{\pm n}$ on both 
sides of \strq , we obtain  $2n$ conditions, which determine
$\bar a_1, ... \bar a_n$.
 %
 
  \vskip 20pt\noindent{\it $\circ$    Example:  sine-Gordon model coupled to gravity 
($t_k = \delta_{k,1}t_n+\delta_{k, -1}\bar  t_{n}$)} 
\smallskip\smallskip

Let us solve, as an example, 
 the string equation for the case  of
 only one pair of nonzero vortex couplings, 
$t= t_1$ and $\bar t=t_{-1}$, which corresponds 
to a  sine-Liouville 
term 
\eqn\sinLiouv{(  t e^{\tilde x R}+
 \bar t  e^{-i \tilde x R})e^{(R-2)\phi}}
added to the world-sheet action  \conepert.  
   In this case  one can retain only $a =a_1$
 and $\bar a = \bar a_{1}$ in \LLbar 
 \eqn\LLbarn{
 \eqalign{
 L&= r   \o   \left( 1+ a /
 \o\right)^{R}\cr 
 \bar L &=    \bar r  \o^{-1} \left( 1+  \bar 
 a\o \right)^{R}}}
  and eq.  \strq\ yields 
\eqn\compasy{  (r \bar r)^{1\over R} 
 \left( 1+a /\o  \right)
  \left( 1+\bar a  \o
\right)  =
\cases{r t      ( \o +  R a )  
  +\mu +  ... & for large $\o$, \cr
\bar r \bar t  (\o ^{-1 }+  R \bar a  )
  +\mu + ... & for small $\o.$}
 }
 Comparing the coefficients in front of  $\o^{0, \pm 1}$
  we get three identities
  \eqn\trieq{(r \bar r)^{1\over R}\bar a =r t     ,
  \ \   (r \bar r)^{1\over R} a
 = \bar r  \bar t ,\ \ 
  (R-1) a\bar a+ \mu   (r \bar r)^{1\over R}=1.}
Taking the limit $\hb\to 0$ limit of  \rrw, we can express the 
  product  $r\bar r$  through the susceptibility
$\chi \equiv  \p_{\mu}^2 \CF $
by
\eqn\rrwpl{
  r\bar r = e^{-\chi}.} 
 From eqs. \trieq\  and \rrwpl\  we get an algebraic
  equation for the susceptibility
\eqn\suscar{\mu e^{{1\over R} \chi}+ \ t \bar t (R-1)
e^{{2-R\over R}\chi}  =1}
and,
 choosing the gauge $r=\bar r$,
 the explicit form of the functions
$z = L(\o)$,
$\bar z=\bar L(\o)$:
\eqn\crfc{\eqalign{
 z&=  e^{-\hf \chi}\ \o \ (1+\bar t\
 e^{{2-R\over 2R}\chi} \ \o^{-1})^R\cr
 \bar z &=  e^{-\hf \chi}\ \o^{-1}\ (1+t\
 e^{{2-R\over 2R}\chi} \ \o)^R.}
 } 
  The two-point correlators   can now be obtained
by plugging $\o(z)$ and $\o(\bar z)$ in \tpca\ and \tpcb\
and expanding in $z$ and $\bar z$.
 To  compute the  one-point correlators, one should    
integrate \muen\ with respect to $\mu$. We will  not do this here, 
since
  the calculation is presented  in  detail  in   \AK .

 In a similar way, if all couplings but  $t_n$ and $t_{-n}$   are zero,   the string
equation is solved by
\eqn\suscarn{\mu e^{{1\over R} \chi}+ (nR-1) \ t_n \bar t_{n}
e^{{2-nR\over R}\chi}  =1}
and
\eqn\crfcn{\eqalign{
 z&=  e^{-\hf \chi}\ \o \ (1+\bar t_{n}\
 e^{{2-nR\over 2R}\chi} \ \o^{-n})^R\cr
 \bar z &=  e^{-\hf \chi}\ \o^{-1}\ (1+t_{n}\
 e^{{2-nR\over 2R}\chi} \ \o^{n})^R.}
 } 
 As expected, the theory compactified at radius $R$ and 
 perturbed by $t_1$ and $t_{-1}$ is equivalent to the theory
  compactified at radius $R/n$ and perturbed by $t_n$ and $t_{-n}$.

    \vskip 20pt\noindent{\it $\circ$    UV$\to$IR flows  in the 
compactified the $c=1$ string theory 
in a vortex background}
\smallskip\smallskip

The critical points in the space of couplings describe  conformal 
matter coupled to gravity. 
 The typical size of the 
world sheet diverges when one approaches the critical point
and is inversely proportional to the cosmological
constant,  defined as the distance to this point. 
At a critical point the    third derivative of 
the free energy in the 
  cosmological constant 
 diverges.

Let us analyze the solution of the string equation in the 
 example considered above, where only the 
three couplings, $\mu, t=t_1$ and $\bar t=\bar t_1$ are nonzero.
In this case the theory   contains
 a single   dimensionless 
parameter
\eqn\lbda{\l\equiv  t\bar t \mu^{R-2}}
which measures the strength of the sine-Liouville 
 perturbation.
The  algebraic equation for the susceptibility 
\suscar\   has three singular points
 \eqn\crcp{
\l = 0,   \ \ \ \ \ 
  \l=\l^*
, \  \ \ \ \l  =\infty}
where we denoted $\l^* =  { (1-R)^{1-R}\over
 (2-R)^{2-R}}$. In the vicinity of these points
 the third derivative of the free energy behaves as
 \eqn\scCF{ \p_{\l}^3\CF \sim \l^{-1}, \ \ 
 \p_{\l }^3\CF \sim (\l-\l^*)^{-1/2},
 \ \ 
 \p_{\l^{-1}}^3\CF \sim (\l^{-1})^{-1}.
 }

 The critical point $\l = 0$  corresponds to  the unperturbed
string theory ($c=1$  compactified  boson  coupled to gravity).
  It was argued in \KKK\ that if  $1< R<2$, the 
  critical point $\l=\infty$ 
 describes another $c=1$ string theory, in which the 
 role of the cosmological constant is played by the
 sine-Liouville coupling. 
 (The latter theory, which can be called sine-Liouville or SL 
 string theory
  is expected to be dual,
 according to the FZZ conjecture\foot{The FZZ conjecture  \FZZ\
  actually  concerns  only  the point $R={3\over 2}$,
which can be considered as a $c=1$ string theory compactified at  $R={3\over 2}$
 and perturbed by 
$t_1\CV_{1}+\bar t_{1}\CV_{-1}$.  },  to the 2D string theory in 
Euclidean black hole background.)  
 Finally, $\l=\l^*$ describes the trivial $c=0$ string theory 
 in which the matter field is the vortex-antivortex plasma 
  with finite correlation 
length.
 
  In this way   the   string theory with a vortex source exists in
two phases,  the ordinary Liouville phase ($0 <\l< \l^*$) 
 and the sine-Liouville  phase ($\l^*<\l< \infty  $). 
In both phases the theory flows to the same  $c=0$    IR fixed point 
$\l=\l^*$.

 Each of the two phases is characterized by its spectrum of local
 operators (the operators creating ``microscopic loops"
 in the world sheet). In the Liouville phase these are the 
 vortex operators of all possible vorticities, which 
 appear in the Laurent expansion of the loop operator $M(z)$.
 The corresponding coupling constants $t_{\pm n}$ 
 are the ``times" of the Toda system.
 
 If we are interested in  the sine-Liouville phase, we should
 be more careful, because the loop operator 
   has two branches   for large  $z$.
 Indeed, let us examine the function 
  $\log \o(z)= \p_\mu S(z)$, which  can be thought of
 as the partition function of a string having a punctured disk
as world sheet,
 with  boundary 
cosmological constant  $z$. 
The  Riemann surfaces of this function  is shown in Fig.1.
    When $1<R<2$, It follows from \LLbarn\  that  
the functions $\log\o(z)$ has two branches at large $z$, 
which are  associated with the 
two UV fixed  points  $\l=0$ and $\l=\infty$
\eqn\twobranches{\o(z)\sim\cases{z, & if  $\l\to 0$\ \  \ \ (branch 
I)\cr
z^{1\over 1-R},& if $\l\to \infty$\ \  (branch II).}
}
In  the SL phase  ($\l^*<\l <\infty$)  
 it is the second   branch   which  is the relevant one.
  In this branch the  partition function on a punctured disk
is expanded as 
\eqn\oSL{
\log\o(z)=  \log \tilde z  + 
\sum_{k\ge 0} c_k \ \tilde z^k, \ \ \  \tilde z = z^{1\over 1 -R}.}
 If we want to calculate the correlation functions of the
microscopic loop operators 
in the SL phase (which are no   vortex operators), 
we can still use   the general  formulas
\tpca \ and \tpcb ,  but we should expand $\o(z)$ in fractional 
powers of $z$.   Near the SL critical point the theory can be again described in terms of a gaussian field $\Phi(\tilde z)$ with  local expansion parameter 
parameter $\tilde z=z^{1\over R-1}$.

\bigskip 

\hskip  60pt
\epsfbox{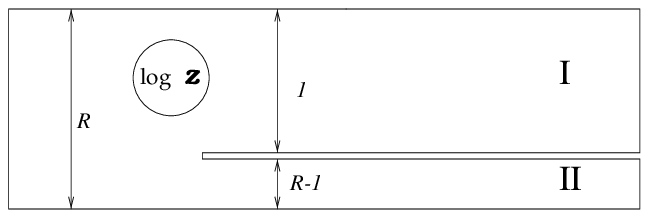}

 \vskip -80pt
 
{\centerline {Fig.1.   The Riemann surface of $
\o(z)$.}}

 \bigskip

 Let us see how the phase transition at $\l=\l^*$ occurs within the description  using  conformal maps. 
The ground state in the dispersionless limit  is described by a closed contour 
$$\g=\{ z(\o); |\o|=1\}$$
in the
$z$-plane, or more correctly  in the Riemann surface of the function 
$\log \o(z)$.
 In terms of the variable $\z= \log z$, this  Riemann surface  is made 
by gluing together  three semi-infinite cylinders
of widths $1$, $R-1$ and $R$.
If we are in the Liouville phase,
then   the contour $\g$ goes around the 
cylinder representing the branch I.  As  $\l$ grows  from 0 to 
$\l^*$, the contour 
moves from infinity towards the branch point. As we will show in a moment, the IR critical point occurs when the contour hits the branch point.   If we are in the SL 
phase, then the contour $\gamma$ wraps the cylinder II and hits the branch point 
when $\l$ approaches $\l^*$ from above.

  We have to show   that    the contour
$\g$  hits the branch point  when the coupling $\l$ reaches the 
critical value  $\l^*$.
 At the branch point $dz/d\o=0$  we have $\o=(R-1)a$.
  On the other hand, the contour $\g$ corresponds to the unit circle $\o=1$.
 The condition $ (R-1)a=1$ is satisfied exactly  at 
 the critical coupling $\l=\l^*$
defined by \crcp .

\vskip 20pt\noindent{\it $\circ$    A description of the IR critical point in terms of the  
dual sine-Gordon theory}
\smallskip\smallskip

An intuitive  physical picture of the flows  in the perturbed theory
can be made  in terms of
the  T-dual 
field $\tilde x$. Below we sketch the scenario proposed by Kutasov
in  \Kutasov  (see also \HK).
If the compactification radius is in the interval ${2\over m+1} < R < {2\over m}$,
then only the vortex operators $\CV_{\pm 1}, ..., \CV_{\pm m}$ are relevant.
 At large  distances on the world sheet the effective couplings grow, 
the  barriers between the minima of the  effective  potential $V(\tilde x)= t_{-m} e^{-m\tilde x/R}+...+t_{m} e^{m\tilde x/R}$ 
become infinite and the original $c=1$ theory splits up  into  to   $m$
theories with  $c=0$.  

In general, the IR matter fields   
need not  be massive.  
     By tuning the coefficients of the potential appropriately one can achieve multicritical behavior $V(\tilde x)- V(\tilde x_0)= C(\tilde x-\tilde x_0)^{2k}+...$, which gives, as was argued in \Kutasov ,  a string theory with central charge
$c= 1- {6\over (k+1)(k+2)}$. 
Therefore one can expect that there  can be multicritical points
that correspond to a direct product of rational $c<1$ 
conformal theories coupled to gravity.
 The general form  of the 
solution \LLbar\  indeed confirms this physical picture.
 For example, if we switch on  all $2m$ couplings
$t_{\pm 1}, ..., t_{\pm m}$ for which the perturbation is relevant, 
then 
we can tune the coefficients in
\LLbar\  and the cosmological constant $\mu$ 
so that the first $m$ derivatives of $z(\o)$ vanish at some point
$z_c$ such that $|\o(z_c) |= 1$.
Then  the IR central charge of the matter field will be
$c_{\rm IR}= 1- {6 \over (m+1)(m+2)}$. 
In this way all unitary $c<1$ string theories   can be obtained by
turning on various perturbations of the 
compactified $c=1$ string theory, as it was suggested in \Kutasov.
  It would be interesting to   exhibit the flow 
 of the  constraint \cexSE\  in Toda hierarchy to the  Douglas  equations
\MikeKDV \ for   generalized  KdV hierarchies.



     \smallskip\smallskip\smallskip
\noindent
{\it Acknowledgments}

The author thanks 
S. Alexandrov,
P. Dorey,
T. Eguchi,
M. Fukuma,
V. Kazakov,
D. Kutasov,
A. Marshakov, 
D. \c Serban, 
P. Wiegmann
and A. Zabrodin 
for stimulating discussions
and useful comments.
This research is supported in part by European network 
EUROGRID HPRN-CT-1999-00161. 
Part of the work is done during the stay  of the author 
at Yukawa Institute, Kyoto.

   \listrefs

\bye